\documentclass[9pt,twocolumn,twoside]{pnas-new}
\templatetype{pnasresearcharticle} % Choose template 
\usepackage{graphicx}
\usepackage{braket}
\usepackage{amsmath}
\usepackage{amssymb}
\usepackage{mathrsfs}
\usepackage{color}

\usepackage[version=3]{mhchem} % Formula subscripts using \ce{}
\usepackage[T1]{fontenc}       % Use modern font encodings
\usepackage{amsmath}
\usepackage{amssymb}
\usepackage{color}
\usepackage[export]{adjustbox}[2011/08/13]
\usepackage{xr}
\usepackage{bm}
\usepackage[normalem]{ulem}
\usepackage{newfloat}
\DeclareFloatingEnvironment[name={Fig. S }]{suppfigure}
\newcommand{\IRS}{I_{\mathrm{R}/\mathrm{S}}}

\newcommand{\w}{\omega}
\newcommand{\Wcm}{\mathrm{W/{cm}^2}}

\title{Real-time probing of chirality during a chemical reaction}

\author[a]{Denitsa Baykusheva}
\author[a]{Daniel Zindel} 
\author[a]{V\`it Svoboda}
\author[a]{Elias Bommeli}
\author[a]{Manuel Ochsner}
\author[a]{Andres Tehlar}
\author[a,1]{Hans Jakob W\"{o}rner}

\affil[a]{Laboratory of Physical Chemistry, ETH Z\"urich, 8093, Zurich, Switzerland}

\leadauthor{Baykusheva}

\significancestatement{Chiral molecules interact and react differently, depending on their handedness (left vs. right). This chiral recognition is the basic principle governing most biomolecular interactions, such as the activity of drugs or our perception of scents. Inspite of this fundamental importance, a real-time (femtosecond) observation of chirality during a chemical reaction has remained out of reach in the gas phase. In the present work, we report this  breakthrough with a seemingly unlikely technique: high-harmonic generation (HHG) in tailored intense near-infrared laser fields. Combining the powerful transient-grating technique with HHG in counter-rotating circularly-polarized laser fields, we follow the temporal evolution of molecular chirality during a chemical reaction from the unexcited electronic ground state through the transition-state region to the final achiral products.}

\authorcontributions{D.B. and H.J.W. designed research; D.B. and A.T. performed research; D.B. analyzed data; D.B. and V.S. performed theoretical calculations; D.Z., E.B. and M.O. contributed new reagents/analytic tools; D.B. and H.J.W. wrote the manuscript.}
\authordeclaration{The authors declare no conflict of interest.}
\correspondingauthor{\textsuperscript{1}To whom correspondence should be addressed. E-mail: hwoerner@ethz.ch}

\begin{abstract}

Chiral molecules interact and react differently with other chiral objects, depending on their handedness. Therefore, it is essential to understand and ultimately control the evolution of molecular chirality during chemical reactions. Although highly sophisticated techniques for the controlled synthesis of chiral molecules have been developed, the observation of chirality on the natural femtosecond time scale of a chemical reaction has so far remained out of reach in the gas phase. Here, we demonstrate a general experimental technique, based on high-harmonic generation in tailored laser fields, and apply it to probe the time evolution of molecular chirality during the photodissociation of 2-iodobutane. These measurements show a change in sign and a pronounced increase in the magnitude of the chiral response over the first 100 fs, followed by its decay within less than 500 fs, revealing the photodissociation to achiral products. The observed time evolution is explained in terms of the variation of the electric and magnetic transition-dipole moments between the lowest electronic states of the cation as a function of the reaction coordinate. These results open the path to investigations of the chirality of molecular reaction pathways, light-induced chirality in chemical processes and the control of molecular chirality through tailored laser pulses.

\end{abstract}

\dates{This manuscript was compiled on \today}
\doi{\url{www.pnas.org/cgi/doi/10.1073/pnas.XXXXXXXXXX}}

\begin{document}

\maketitle
\thispagestyle{firststyle}
\ifthenelse{\boolean{shortarticle}}{\ifthenelse{\boolean{singlecolumn}}{\abscontentformatted}{\abscontent}}{}

%\section{Introduction}
 
\dropcap{T}he two enantiomers of a chiral molecule interact differently with chiral receptors and with chiral light. The former effect is the basis of chiral recognition, an essential mechanism of biomolecular function. The latter is the principle of optical techniques for detecting chirality. Although extensive control over molecular chirality has been achieved in enantio-selective synthesis of molecules \cite{frantz00a,yoon03a,quasdorf14a}, chiral sensitivity has been lacking from all femtosecond time-resolved probes of chemical reactions in the gas phase demonstrated to date. The main origin of this shortcoming is the weakness of chiral light-matter interactions, which rely on magnetic-dipole, electric-quadrupole and higher-order interactions. For this reason, circular dichroisms (CD) in absorption spectroscopies are typically very weak effects in the range of 10$^{-6}$ to 10$^{-3}$ relative signal changes. Nevertheless, time-resolved circular dichroism (TRCD) techniques have gained increasing attention over the last 15 years \cite{Meyer-Ilse2013}, and nowadays cover the IR~\cite{Bonmarin2008,Rhee2009}, visible~\cite{Niezborala2006}, and UV~\cite{Abramavicius2005, Hache2009,meyer-ilse12a,oppermann19a} spectral regions. The promise of extending such measurements to the X-ray domain has recently been predicted \cite{rouxel17a,zhang17a,rouxel19a}. Unfortunately, the inherent weakness of the optical CD effects and the related technical challenges have so far limited TRCD measurements to the condensed phase and to picosecond or longer time scales.

The development of more sensitive methods, applicable to isolated molecules in the gas phase has therefore received considerable attention \cite{quack89a}. Important developments include microwave three-wave-mixing spectroscopy~\cite{Patterson2013}, Coulomb explosion imaging (CEI,~\cite{Herwig2013, Pitzer2013}), laser-induced mass spectrometry~\cite{Li2006, Bornschlegl2007}, photoelectron circular dichroism (PECD, \cite{Ritchie1975,Boewering2001,Powis2008, Ferre2014, Beaulieu2016}), and high-harmonic generation (HHG) in elliptically polarized \cite{Cireasa2015} and two-color laser fields \cite{baykusheva18a}. 
%In this respect, a noteworthy approach towards recording chiral dynamics was proposed and demonstrated in a proof-of-principle experiment by Kitamura \textit{et al.}~\cite{Kitamura2001}, whereby chiral asymmetries in an initially achiral system (CD$_4$) undergoing zero-point vibrations were observed. In a direct pump-probe CEI experiment, Hansen \textit{et al.}~\cite{Hansen2012} studied the time evolution of the laser-induced torsional motion of a chiral biphenyl molecule. However, none of these studies did provide access to an observable quantifying the time-dependent chiral response. 
%An even more challenging task is the investigation of chiral single-molecule response on an ultrafast timescale, where probing the relevant dynamics typically requires the application of non-linear or strong-field time-resolved imaging techniques. Despite being one of the most mature techniques for studying chiral phenomena in the strong-field regime, 
Time-resolved PECD (TR-PECD) has so far only been demonstrated in the single-photon-ionization regime using laser pulses in the visible domain \cite{Comby2016}. As a consequence, it has remained restricted to probing dynamics in highly excited states rather than photochemical reactions. The extension of TR-PECD to XUV and soft-X-ray radiation derived from high-harmonic~\cite{Svoboda2019} and XFEL~\cite{Facciala2019, Schmidt2019} facilities currently represents an active research direction.
%photophysical process on the example of the ultrafast relaxation of the 3s Rydberg state of fenchone. Photoexcitation-induced circular dichroism (PX(E)CD)~\cite{Beaulieu2018a}, a novel extension of PECD that is sensitive to the coherent helical motion of bound-state electrons, was used to follow the dynamics of Rydberg wavepackets in camphor and fenchone and the associated chiral vibrational motion. By elegantly combining above-threshold ionization and PECD in an interferometric two-color pump-probe, Beaulieu \textit{et al.}~\cite{Beaulieu2017} were able to resolve the few-attosecond time delays between electrons emitted in the forward and the backward directions upon photoionization of a chiral molecule. On a different note, chirality-sensitive high-harmonic spectroscopy (cHHG)~\cite{Cireasa2015}, a recently introduced all-optical technique, offers an indirect access to sub-fs chiral electron dynamics via the sub-cycle mechanism underlying the high-harmonic generation (HHG) process~\cite{Smirnova2015}.

Here, we introduce an all-optical, ultrafast and nearly general experimental technique for probing the time-dependent chirality of molecules undergoing a photochemical reaction. This method combines the established sensitivity of HHG spectroscopy to photochemical dynamics \cite{woerner10b,woerner11c,kraus12a} with the unique chiral sensitivity of HHG in two-color counter-rotating circularly-polarized (so-called ''bi-circular'') laser fields, which was theoretically predicted \cite{smirnova15a,Ayuso2018a,Ayuso2018b} and experimentally demonstrated on unexcited molecules very recently \cite{baykusheva18a,Fleischer2017,harada18a}.

The key advantages of our pump-probe technique are its sensitivity to gas-phase samples, its temporal resolution and the background-free detection of the excited-state dynamics. We demonstrate the ability of the technique to follow in real time the evolution of molecular chirality along a photochemical reaction pathway. This development paves the way to detecting the chiral interactions that underlie enantio-specific chemical reactions. Building on the self-probing capabilities of high-harmonic spectroscopy \cite{baker06a,smirnova09a}, our technique also provides access to the attosecond electron dynamics \cite{kraus15a} of the underlying chiral response \cite{baykusheva18a} and its dependence on the chemical reaction pathway. With the recent extensions of high-harmonic spectroscopy to solids \cite{ghimire11a,luu15a,vampa15a} and liquids \cite{Luu2018a}, our technique will rapidly become applicable to all phases of matter. In the liquid phase, it will give access to the interactions underlying chiral recognition. In the solid state, it might open new approaches to elucidate dynamics in chiral crystals, which include skyrmions in chiral magnets, unconventional pairing in chiral superconductors, as well as unique magnetoelectric effects in chiral metals. Its all-optical nature paired with its high degree of chiral discrimination will facilitate its application to dense or highly-charged forms of matter.

\section*{Results}

\begin{figure*}[b!]
  \includegraphics[width=1.\textwidth,center]{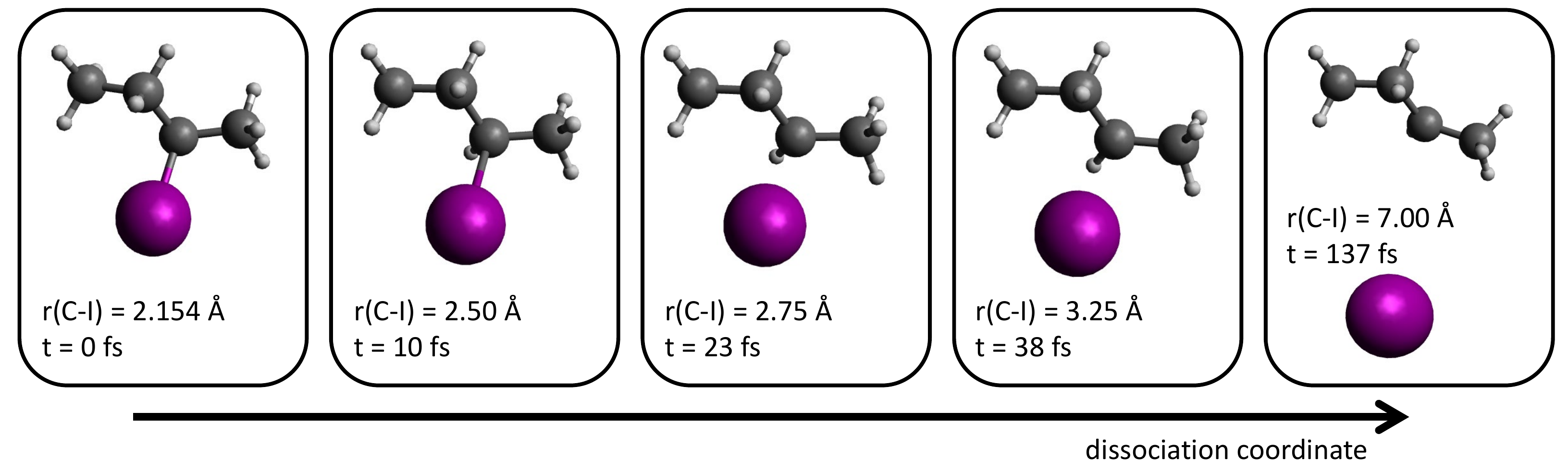}
  \caption{Illustration of the investigated photochemical reaction. An initially chiral molecule is photoexcited by a femtosecond laser pulse centered at 266~nm. The ensuing photodissociation reaction leads to a time-dependent change in the structure of the chiral center. The left-most panel displays the equilibrium geometry of $(R)$-2-iodobutane (s. SM for details). The structures in the other panels have been obtained by varying the C-I bond distance and relaxing all remaining degrees of freedom. The corresponding time delays have been taken from the wave packet calculations shown in Fig. S6.}
  \label{fig:scheme}
\end{figure*}

The light-induced dissociation of 2-iodobutane ($2-\mathrm{C_4H_9I}$), depicted in Fig.~\ref{fig:scheme}, was selected as an illustrative chemical reaction. Photoexcitation of this molecule at 266~nm accesses the $\tilde{\rm A}$-band and mainly populates the ${}^3Q_0^+$-state in a parallel transition, as revealed by magnetic circular-dichroism measurements \cite{Gedanken1987}. The repulsive potential-energy curves of the ${}^3Q_0^+$ and the two close-lying electronic states (${}^3Q_1$ and ${}^1Q_1$) are given in the Supplementary Information (SI) Appendix, Fig. S3. Following excitation to the ${}^3Q_0^+$-state, the molecule dissociates into a 2-butyl radical and a spin-orbit-excited iodine atom ($\mathrm{I}^\ast$, ${}^2P_{1/2}$) on an ultrafast time scale.

The breakage of the C-I bond also induces a major modification of the chiral structure of the molecule. The main idea of our work is to map out the temporal evolution of the molecular chirality as the laser-induced photodissociation takes its course. We thereby answer the following questions. Is the 2-butyl radical produced in a chiral or achiral state? How fast do the signatures of the changing chirality evolve from the reactant to the product state? Which properties of molecular chirality is our method sensitive to?

\begin{figure*}
  \includegraphics[scale=0.5,center]{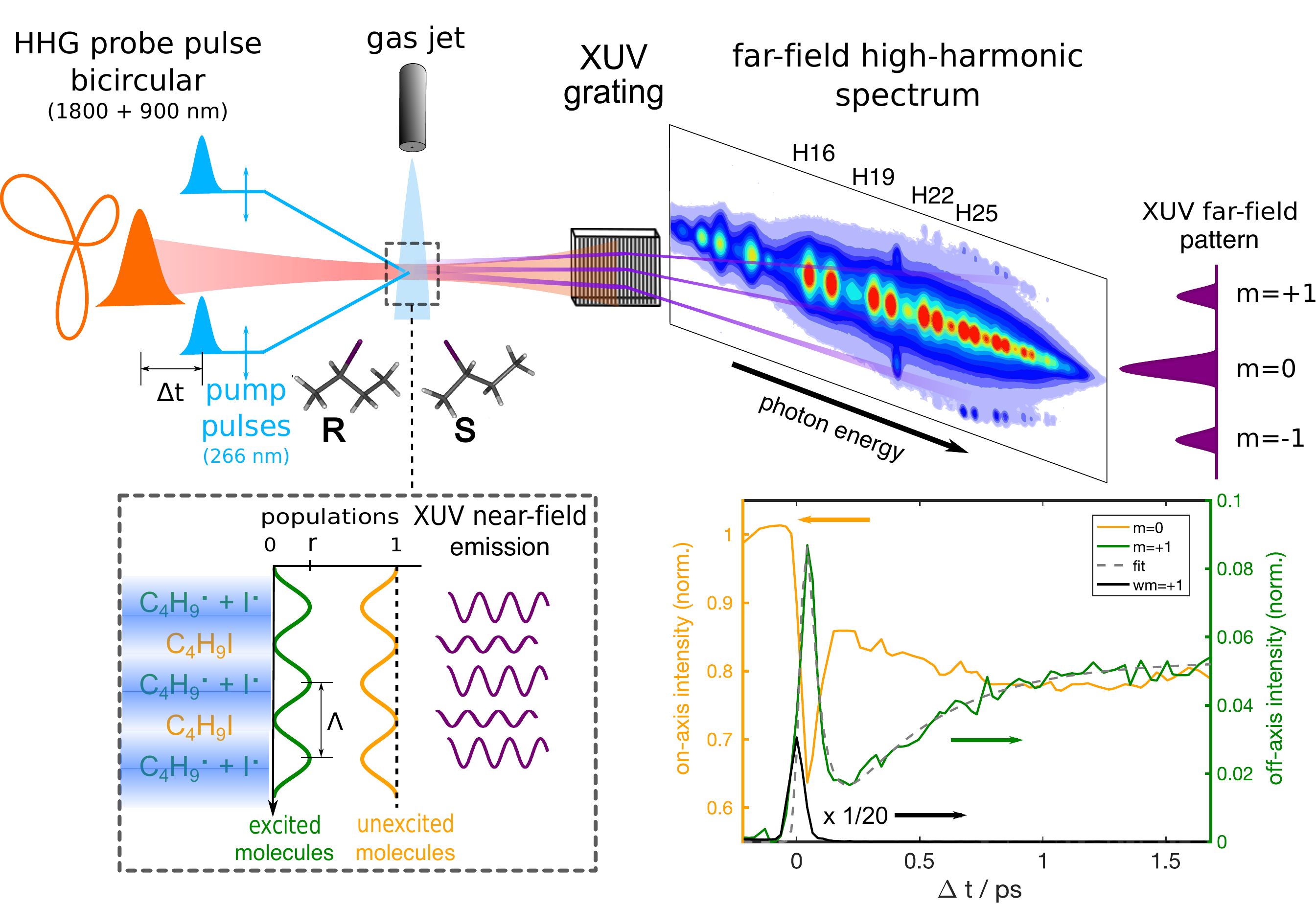}
  \caption{Principle of bi-circular high-harmonic transient-grating spectroscopy. Two noncollinear pump beams are crossed under a small angle, creating an intensity grating which spatially modulates the molecular excitation. High-harmonic radiation is generated by a temporally delayed bi-circular field consisting of a 1800 nm pulse and its second harmonic centered at 900 nm. In the far-field, the radiation diffracted by the optical grating is spatially separated from the undiffracted light. The panel in the bottom right corner shows the time-dependent signals observed in 2-iodobutane after excitation at 266~nm followed by HHG driven by a bi-circular 1800+900~nm field observed in the undiffracted orders (orange curve, left-hand vertical axis) and averaged $wm=\pm1$- (black) and $m = \pm1$- (dark green) orders (left-hand axis) for harmonic 23 of 1800~nm ($\sim 15.8$~eV). The origin of the time-scale has been set to the maximum of the wave-mixing signal.}
  \label{fig:setup}
\end{figure*}

The experimental setup is illustrated in Fig.~\ref{fig:setup}. A transient grating (TG,~\cite{mairesse08a,woerner10b,woerner11c}) is produced by crossing two linearly-polarized vertically-offset 266~nm pulses in a supersonic expansion of gaseous 2-iodobutane. A periodic intensity variation pattern is produced in the focal plane, and, as efficient photoexcitation is restricted to the high-intensity regions, this  modulation translates into alternating planes of excited and ground-state molecules. The TG is probed by a bi-circular field consisting of the superposition of two circularly polarized $\sim$40~fs pulses centered at 1800 and 900~nm with opposite helicities. The corresponding electric field describes a Lissajous figure resembling a clover leaf, as illustrated on the left-hand side of Fig.~\ref{fig:setup}.
As the HHG radiation emitted from the individual layers of the grating in the near field is characterized by different amplitudes and phases, a diffraction pattern is formed in the far field, leading to a background-free detection of the excitation (s. Fig.~S1). As revealed by the data presented in the SI Appendix, the typically achieved diffraction efficiencies reach up to $7-8 \%$.

%\begin{figure}
%  \includegraphics[scale=0.5,center]{figures-work/first-draft-pdfs/fitres_H23_mp1_raw.pdf}
%  \caption{Time-dependent transient-grating signal observed in 2-iodobutane after excitation at 266~nm followed by HHG driven by a bi-circular 1800+900~nm field. The displayed signals correspond to the time evolution of the harmonic signal in the undiffracted orders (black curve, left-hand vertical axis) and averaged $wm=\pm1$- (blue) and $m = \pm1$- (red) orders (left-hand axis) for
%harmonic 23 of 1800~nm ($\sim 15.8$~eV). The origin of the time-scale has been set to the maximum of the wave-mixing signal.}
%  \label{fig4}
%\end{figure}

The experiment consisted in measuring the intensity of the diffracted and undiffracted high-harmonic emission as a function of the time delay between the excitation and probe pulses. The dynamics of the 2-iodobutane photodissociation were investigated for pump-probe delays $\Delta t$ spanning the range from -0.15 to 5~ps. The panel in the bottom right corner of Fig.~\ref{fig:setup} shows the short-time evolution of signal in harmonic order 23 (of 1800~nm, $\sim 15.8$~eV) recorded in racemic 2-iodobutane. The orange curve represents the undiffracted intensity ($I_{m=0}$), while the averaged diffracted (${m=\pm1}$) signals are shown in dark green. Temporal overlap of the pump and probe pulses leads to high-order wave mixing \cite{bertrand11a}, which appears spatially separated from the other signals on the detector (see SI Appendix, Fig.~S1). This wave-mixing signal, averaged over its two components ($wm=\pm1$) is shown as black line. It accurately determines the characteristic cross-correlation time of the experiment ($\approx45$~fs).

The temporal evolution of the undiffracted intensities is qualitatively similar for all harmonic orders, as is the time evolution of all observed diffracted intensities. We characterize the photodissociation dynamics observed in each harmonic order by quantifying the exponential time scale of intensity build up in the diffracted signals $\tau_\mathrm{rise}$, defined by the phenomenological kinetic expressions given in Eqs.~(1-2) in the SI Appendix. The extracted values of $\tau_\mathrm{rise}$, averaged over multiple measurements, are listed in the SI Appendix (s. Tab.~I). Notably, the observed (averaged) dissociation time scale ($\tau_\mathrm{rise}\approx370$~fs) is longer than the estimations from wave-packet calculations (s. Sec.~IV in the SI Appendix) or extrapolations from previous studies (s. discussion in Ref.~\cite{Corrales2014}). This observation is in line with the conclusions from previous TG-HHG studies on the photodissociation dynamics of $\mathrm{Br}_2$~\cite{Woerner2010a} and other alkyl halides employing linearly polarized driving fields~\cite{Tehlar2013b}.

%Table~\ref{tab1} lists the extracted values of $\tau_\mathrm{rise}$, averaged over multiple measurements, as a function of harmonic energy. A trend towards increasing $\tau_\mathrm{rise}$-values as  a function of the harmonic order is discernible. Both observations are in line with the conclusions from  previous TG-HHG studies on the photodissociation dynamics of alkyl halide employing linearly polarized fields.

We now turn to the characterization of the time-dependent chirality during the photodissociation. Enantiomerically-enriched samples ($ee\sim 60-65 \%$) of $(R)$- and $(S)$-2-iodobutane were synthesized in house as described in the SI Appendix (Section II). The chirality-sensitive measurements were performed by monitoring the signal of the undiffracted emission ($I_{m=0}$) as the ellipticity of the $\w$ and $2\w$ laser fields constituting the bi-circular field is varied by rotating the main axis of the achromatic quarter-waveplate (QWP) from $\alpha=30^\circ$ to $\alpha=150^\circ$. In this manner, the polarization of the $\w$ and $2\w$ laser fields is changed from circular via elliptical to linear and then back to circular again. At both $\alpha=45^\circ$ and $\alpha=135^\circ$, the two pulses combine into a bicircular field, but with opposite helicities. These two helicities interact differently with the chiral molecules, resulting in a CD effect in the intensity of the emitted high-harmonic radiation. Expressions~(III-IV) of the SI Appendix outline the procedure for extracting the frequency- and ellipticity-dependent circular dichroism $CD(n\w, \alpha)$ from the normalized HHG intensities  $\IRS(n\w)$ recorded from each enantiomer, where $n\w$ denotes the frequency of the $n^\mathrm{th}$-harmonic of the fundamental frequency $\w$. We further define an ellipticity-averaged quantitative measure for the observed CD by averaging the ellipticity-resolved $CD(n\w,\alpha)$ in the region close to the two opposite bi-circular configurations ($\alpha\in [35^\circ-55^\circ]$ resp. $\alpha\in [125^\circ-145^\circ]$):

\begin{align*}
\overline{CD}^\pm(n\w) = \int_{\alpha_{\rm min}}^{\alpha_{\rm max}} CD(n\w,\alpha)\mathrm{d}\alpha,\numberthis \label{eq:CDpm_def}
\end{align*}
whereby the superscript $\pm$ specifies the two possible helicity combinations of the $\w$ and $2\w$ laser fields. This procedure minimizes the asymmetries resulting from imperfections of the optical components and the optical path misalignments and expresses the CD as a function of the photon energy only. 

\begin{figure*}
  \includegraphics[width=1.\textwidth,center]{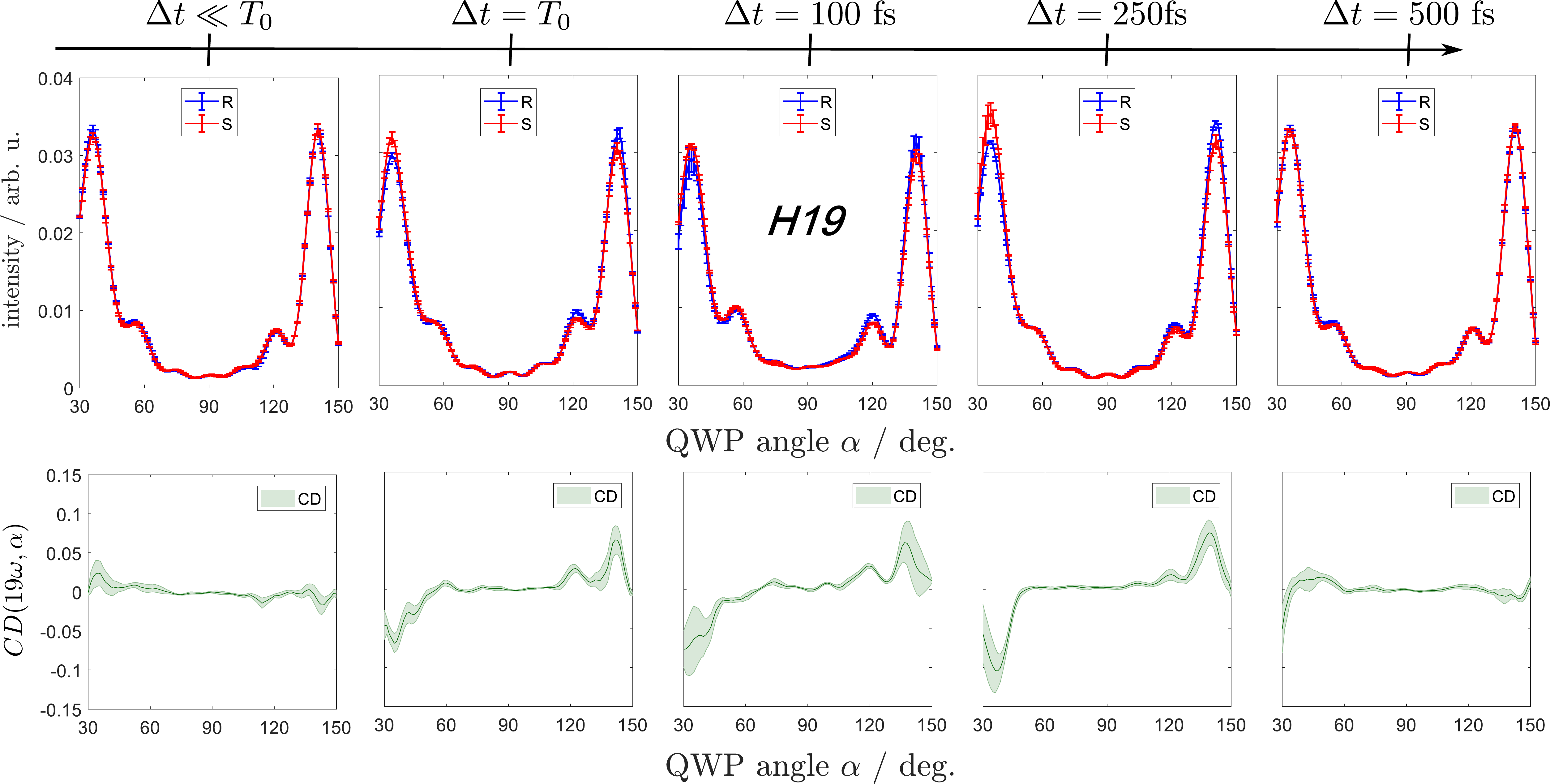}
  \caption{\textit{Upper row}: Normalized responses of the $(R)$- (blue) and the $(S)$- (red) enantiomers of 2-iodobutane at different pump-probe delays as a function of ellipticity expressed in terms of the QWP rotation angle $\alpha$. The signals correspond to harmonic 19 of 1800~nm, or $\sim13.1$~eV. \textit{Bottom row}: Ellipticity-dependent circular dichroism at harmonic 19 for each time step.}
  \label{fig:CD_H19}
\end{figure*}

\begin{figure*}
  \includegraphics[width=1.\textwidth,center]{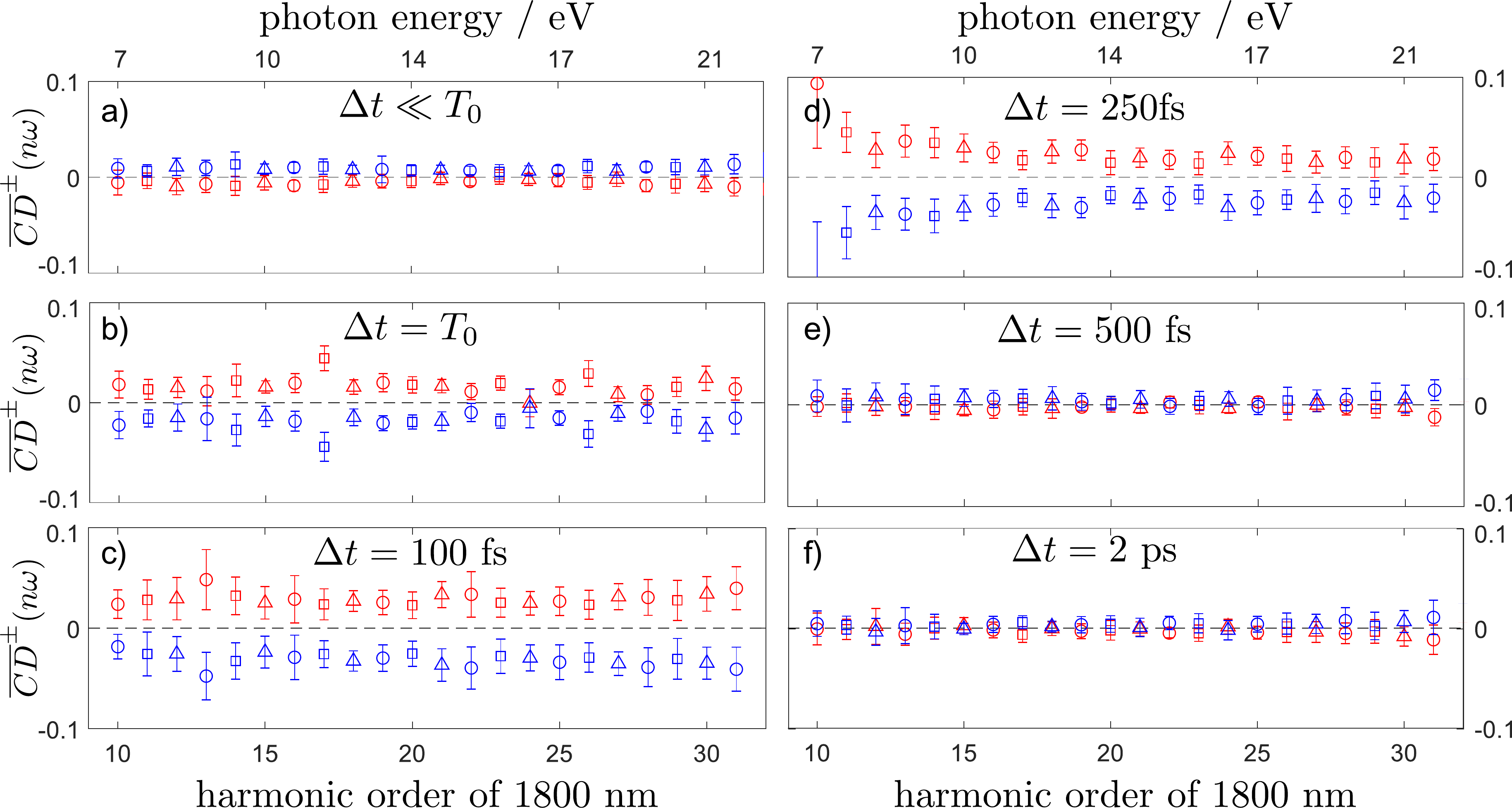}
  \caption{Ellipticity-averaged circular dichroism observed in 2-iodobutane as a function of the photon energy. The data in each panel is recorded at a different value of the pump-probe delay $\Delta t$.}
  \label{fig:CD_2D}
\end{figure*}

Figure~\ref{fig:CD_H19} shows the high-harmonic signal in harmonic order 19 (H19) in terms of the undiffracted-signal intensities emitted from the individual enantiomeric samples (\textit{(R)} and \textit{(S)}), as blue and red curves in the upper row. The differential ellipticity-resolved CDs are shown in the lower row. Since the two employed samples have similar enantiomeric excesses of 60-65~$\%$, the reported CDs are lower than those of the enantiopure samples by approximately this amount. The antisymmetry of the green curves in Fig.~\ref{fig:CD_H19} shows that a reversal of the rotation direction of the driving fields has the same effect as changing the handedness of the molecular sample, eliminating the possibility of artefactual contributions.
The temporal evolution of $\overline{CD}^\pm(n\omega)$ for all harmonic orders is compactly presented in Fig.~\ref{fig:CD_2D}, where the individual panels correspond to the points of the pump-probe delay grid.

The CD was first measured in the absence of the pump pulses to characterize the chiral response of the unexcited molecules. The obtained CD is on the order of 2$\%$ at H19 (see Fig.~\ref{fig:CD_H19}, left panel) and across the whole range of harmonic orders (see Fig.~\ref{fig:CD_2D}a). These values are notably smaller than in the case of methyloxirane \cite{baykusheva18a} and limonene \cite{harada18a}. Although the CDs are somewhat diminished by the limited enantiopurity of our sample (see above), the CDs scale approximately linearly with the enantiomeric excess, such that the smallness of the CD must be a property of 2-iodobutane in its electronic ground state.

In the presence of the pump pulses, the CD changes sign and significantly increases in magnitude (Fig.~\ref{fig:CD_H19}) across all detected high-harmonic orders (H10-H31, Fig.~\ref{fig:CD_2D}b). At and after time zero ($T_0$), the maximal CD values in H19 amount to $6-7 \%$. The averaged $|\overline{CD}^\pm(19\w)|$ value amounts to $\sim 2\%$ at $T_0$, increases to $\sim2.5 \%$ after 100~fs, and eventually reaches a maximum of $2.7 \%$ at 250~fs. After 500 fs have elapsed, there is essentially no detectable chiral response left and the same is true for the subsequent points on the time grid. Other harmonic orders exhibit a qualitatively similar behaviour, although there are differences concerning the time delay maximizing the chiral response (see, for instance, H10 and H13 in Fig.~\ref{fig:CD_2D}, for which $|\overline{CD}|$ maximize at $\Delta t = 250$~fs and $100$~fs, respectively).

The results summarized in Fig.~\ref{fig:CD_2D} lead to the following main conclusions: \textit{i)} a strong chiroptical response appears during the excitation by the short 266~nm pulse, which is opposite in sign to that of the unexcited molecules; \textit{ii)} the induced chiral response persists up to $\gtrsim 250$~fs, after which it decays, signalling the formation of an achiral product state; \textit{iii)} the chiral response observed below the ionization energy of 2-iodobutane (9.13~eV) displays a strong dependence on the emitted photon energy.

\section*{Discussion}

In this section, we discuss the experimental observations and provide a qualitative interpretation. There are basically two possible mechanisms for explaining a pump-induced variation of the CD. First, we discuss the role of the rotational dynamics induced by photoexcitation, and then proceed with an analysis of the transient CD changes associated with the modifications of the chiral structure of the molecule. 

Resonant one-photon-excitation leads to the selective excitation of a sub-ensemble of molecules that have their transition-dipole moments aligned along the polarization direction of the pump pulse(s). This effect leads to a partial alignment of the excited molecules that decays on the picosecond time scale. Since CD effects in bi-circular HHG strongly depend on the orientation of the molecule relative to the laser field \cite{Ayuso2018a}, partial alignment can enhance the CDs and the subsequent rotational dephasing could explain their decay. Therefore, we have performed detailed calculations on the time dependence of the alignment of 2-iodobutane caused by the pump pulses. These calculations, that are summarized in the SI Appendix (section VI and Fig.~S8), show that for typical expected rotational temperatures of 5-30~K in our supersonic expansion, the rotational anisotropy decays on a time scale of 11-5~ps. Since these time scales are much longer than the observed CD dynamics, we can exclude rotational dynamics as the origin of our observations.

Therefore, the observed CD dynamics are most likely to be caused by the structural modification of the chiral center as the reaction progresses. In the following, we discuss a theoretical model that captures the essential features of the observed chiral dynamics. Our model builds on previous theoretical analysis of bi-circular HHG from chiral molecules \cite{Cireasa2015,Smirnova2015, baykusheva18a}, which has linked the induced chiral response with the interplay of electric- and magnetic-dipole transitions occurring during the propagation of the electron in the continuum. If several valence shells lie sufficiently close in energy ($\lesssim 2$~eV), strong-field ionization can leave the cation in a coherent superposition of several  electronic states. During propagation of the electron in the continuum, laser-induced couplings between these states lead to the emergence of HHG ''cross-channels'', whereby the returning electron recombines with the cation in a state that is different from the one created during ionization. Circular dichroism emerges through the participation of the weak magnetic-field component of the driving laser pulses in the laser-induced dynamics, whereby the chiral sensitivity is exclusively conveyed by the cross channels mentioned above. Our previous analysis has shown that although the corresponding changes in the induced dynamics are extremely small ($10^{-4}$ level in the electron-hole density), a substantial CD (up to 13$\%$ in the case of methyloxirane) can result \cite{baykusheva18a}.

Limiting our analysis of the $\tilde{\rm A}$-band dynamics to photoexcitation via the ${}^3Q_0^+$-state as explained above, we first extract the evolution of the expectation value of the C-I-bond distance ($\langle r_\mathrm{CI} \rangle (\Delta t)$) as a function of the pump-probe delay $\Delta t$ from a wavepacket propagation on the corresponding one-dimensional potential-energy curve (see SI Appendix, Sec.~IV). 
Within a classical-trajectory treatment, the bond length is bijectively mapped onto the pump-probe delay, which allows us to express all relevant potential-energy curves and coordinate-dependent transition moments as a function of the pump-probe delay. Detailed information on the quantum-chemistry calculations is given in the SI Appendix (Sec.~III). 
%Its photoelectron spectrum~\cite{Dromey1974,KimuraBook} consists of two narrow lines (at 9.13 and 9.68~eV) separated by the spin-orbit (SO) interaction, followed by a series of broad overlapping bands starting at $\sim 11$~eV. Considering the pulse intensities employed in the experiment ($\lesssim 5\times10^{13} \Wcm$ and wavelengths of 1800+900~nm), strong-field  ionization is mainly restricted to the removal of an electron from the highest-occupied molecular orbital (HOMO) via tunnelling.

Considering the pulse intensities employed in the experiment ($\lesssim 5\times10^{13} \Wcm$ and wavelengths of 1800/900~nm), strong-field  ionization from the ${}^3Q_0^+$-state populates mainly the ground ($\tilde{X}^+$, split into two spin-orbit components with vertical ionization potentials $I_\mathrm{p}$ of $\approx 9.13$~eV and $\approx 9.68$~eV) and the first excited ($\tilde{A}^+$, $I_\mathrm{p}\approx11.08$~eV) states~\cite{Dromey1974,KimuraBook}. Neglecting the spin-orbit interaction in the cationic states, this gives rise to in total four channels contributing to the HHG process. Two channels are formed by strong-field ionization to the $\tilde{X}^+$ or $\tilde{A}^+$ state of the cation and photorecombination of the continuum electron with the same state. These channels are accordingly labeled $XX$ or $AA$. In addition, there are two ''cross channels'' ($XA$, $AX$). Under these premises, the treatment of the laser-induced dynamics reduces to a system of two coupled ordinary differential equations that has a straightforward solution, as presented in Section~V of the SI Appendix. After integrating Eq.~(5) of the SI Appendix over the duration $\overline{\tau}^\Omega$ of a given electron trajectory (typical values ranging from 1.4 to 1.8~fs for our experimental conditions), the calculated coefficients $\{c_{IJ}(\Delta t)\}$ (with $\{I,J\}\in\{X,A\}$) for each channel are used to obtain the frequency-domain HHG dipole response (at a photon energy $\Omega$) as a function of the pump-probe delay~\cite{Cireasa2015}:
%\begin{figure*}[h!]
\begin{align*}
\mathbf{d}(\Omega, \Delta t) = \sum_{IJ} c_{IJ}(\Delta t) a_{ion}^I(\Omega,\Delta t)\left(\frac{2\pi}{i\overline{\tau}^\Omega_{IJ}(\Delta t)}\right)^{(3/2)}e^{iS(\Omega,\overline{\tau}_{IJ}^\Omega)}\mathbf{a}_{rec}^J(\Delta t),\numberthis \label{eq:dHHG}
\end{align*}
%\end{figure*}
where $S(\Omega, \overline{\tau}_{IJ}^{\Omega})$ corresponds to the semi-classical action. 

%
%The ground electronic state of the cation $\tilde{X}^+$, split by the SO interaction into two doublets separated by XX~eV at  equilibrium distance, and the first-excited $\tilde{A}^+$ state are separated by $\sim 1.5$~eV. The higher-lying $\tilde{B}^+$- and $\tilde{C}^+$-states start from XX and XX~eV, respectively (corresponding to vertical $I_\mathrm{p}$-s of XX and XX~eV), are neglected in our treatment of the continuum laser-induced dynamics. 

%
\begin{figure}
  \includegraphics[scale=0.5,center]{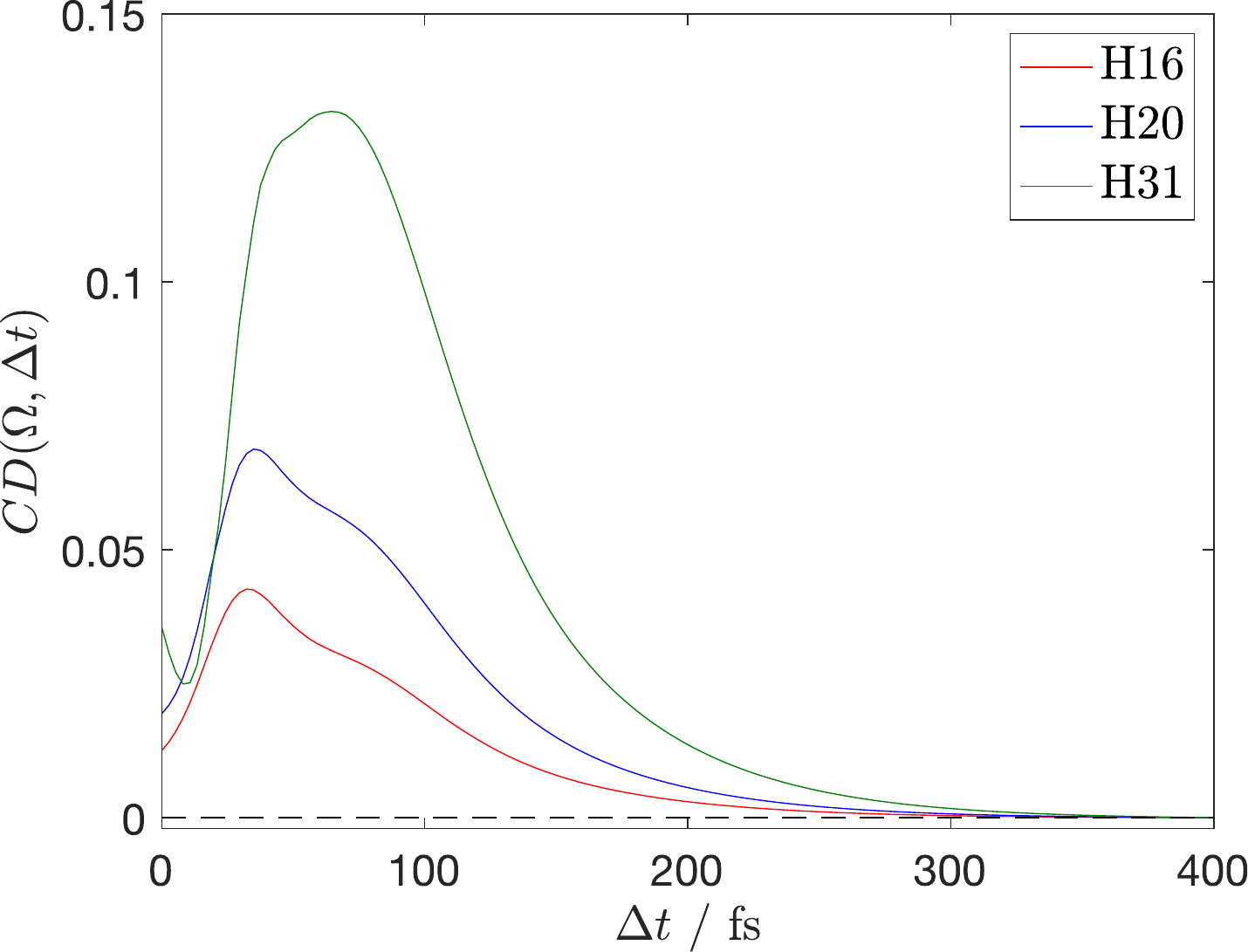}
  \caption{Time-dependent CD signal, evaluated according to Eq.~(\ref{eq:CD_t}) at photon energies corresponding to $H16$, $H20$, and $H31$ of 1800~nm. The instantaneous direction of the ionizing bi-circular field is set parallel to the C-I-bond axis. The signal is averaged over the azimuthal angle associated with this axis.}
  \label{fig:theory}
\end{figure}

Here, we are primarily interested in rationalizing the temporal profile of the chiral response as a complete quantitative description of the HHG process for many-electron systems is still out of reach. Therefore, we set the ionization ($a_{ion}(\Omega, \Delta t)$) and the recombination ($\mathbf{a}_{rec}(\Omega,\Delta t)$) matrix elements in Eq.~(\ref{eq:dHHG}) to unity, which amounts to neglecting their dependence on the molecular-frame ionization/recombination directions. With these approximations, the chiral response is entirely encoded in the variation of the electric- and magnetic-dipole matrix elements within the sub-cycle temporal interval between ionization and recombination. The time-dependent circular dichroism $CD(\Omega,\Delta t)$ is estimated by calculating the spectral HHG intensity $\tilde{S}(\Omega; \Delta t)\propto |\mathbf{d}(\Omega, \Delta t)|^2$ for a $(R)$-2-iodobutane molecule with its C-I-bond-axis aligned along the instantaneous electric-field direction as well as for its $(S)$-enantiomer and subsequently forming the differential response:
\begin{align*}
CD(\Omega,\Delta t) = 2\frac{\tilde{S}^{(R)}(\Omega,\Delta t)-\tilde{S}^{(S)}(\Omega,\Delta t)}{\mathrm{max}\left(\tilde{S}^{(R)}(\Omega,\Delta t)+\tilde{S}^{(S)}(\Omega,\Delta t)\right)}.\numberthis\label{eq:CD_t}
\end{align*}
Figure~\ref{fig:theory} shows the CD obtained with the above relation for three selected harmonic orders ($H16$, $H20$ and $H31$) as a function of the pump-probe delay. Since our model does not contain the structure-sensitive contributions from the ionization and the recombination steps, a quantitative prediction of the magnitude of $\overline{CD}^\pm(\Omega)$ is not expected. 

However, a special feature of our HHG scheme of probing chiral dynamics is the fact that the chiroptical response entirely originates from the second, continuum propagation step, in which the electron can be treated as decoupled from the parent ion in good approximation. Therefore, the time dependence of the chiral response will mainly be dictated by the variation of the chiral structure of the probed molecule, which is encoded in the geometry-dependent electric and magnetic dipole-transition-matrix elements of the cation. In our work, the latter are calculated with high accuracy using modern quantum-chemistry methods (SI Appendix, Sec. V). 

Our results predict a rapid increase of the CD signal at early time delays, followed by one relatively sharp local maximum around 50~fs and a broad shoulder feature around 100~fs, and then monotonous decay by 2-3 orders of magnitude within $\sim250-300$~fs. The overall dynamics of the calculated CDs therefore capture the main features of the experimental results, which show local maxima around 100~fs and a subsequent decay.  Moreover, the results in Fig.~\ref{fig:theory} predict a pronounced spectral dependence of the time-dependent chiral response, which is also present in the experimental results shown in Fig.~\ref{fig:CD_2D}. This comparison shows that our simple model captures the salient features of the measured chiral dynamics. Our calculations further allow us to conclude that the decay of the chiral response and its vanishing for long  delays reflect the progressive loss of chirality during the C-I bond breaking process. We have carried out a detailed comparison of the predictions of our theoretical model including or neglecting the planarization of the 2-butyl radical (s. section III~B in the SI Appendix). This comparison has shown that the geometric relaxation of the 2-butyl radical does not qualitatively change the predicted CD dynamics, which are therefore dominated by the detachment of the iodine atom from the molecular backbone. This conclusion is further supported by our analysis of the time scale of the interconversion dynamics following the initial formation of the chiral enantiomeric 2-butyl radical conformer. The calculations presented in Section VIII of the SI Appendix corroborate the fact that in the absence of coherent stereomutation, rapid racemization of the initially formed chiral 2-butyl radical will take place.

Turning to the remaining discrepancies, we note that the measured CDs appear to decay more slowly than the calculated ones. Similarly, we noted above that the (non-chiral) signals suggest a longer dissociation time (340~fs) than predicted by our one-dimensional wave-packet calculations. The consistent deviation of these two classes of observables from their calculated counterparts most likely originate from neglecting the structure-dependent ionization and recombination matrix elements, because all other quantities have been accurately incorporated into our theoretical model. 

\section*{Conclusion}

In this article, we have introduced a new technique for detecting time-dependent chirality during a chemical reaction. We have applied this technique to study the chirality changes occurring in the course of the photodissociaton of 2-iodobutane. With the aid of a conceptually simple theoretical treatment based on high-level {\it ab-initio} quantum-chemical calculations, we have been able to correlate the time evolution of the CD signal with the coordinate dependence of the electric- and magnetic-dipole matrix elements that encode the chiral response in the HHG process.

This technique is sufficiently sensitive to be applicable to the gas phase, but is equally applicable to liquids \cite{Luu2018a} and solids \cite{ghimire11a} in future experiments. Its relatively large CD effects, comparable in magnitude to PECD, combined with its all-optical nature make it a powerful chiral-sensitive method for detecting ultrafast changes in molecular chirality. The ability to probe chiral photochemical processes on such time scales opens up a variety of  possibilities for investigating chiral-recognition phenomena, such as the processes that determine the outcome of enantioselective chemical reactions. Paired with recent developments in condensed-phase high-harmonic spectroscopy, our technique will rapidly be applied to all phases of matter, where it will unlock the study of a range of chiral phenomena on ultrashort time scales.

\section*{Methods}

Details on the experimental setup and data evaluation are given in SI Appendix I. The synthesis of enantiomerically-enriched 2-iodobutane is described in SI Appendix II. The theoretical work including the quantum-chemical {\it ab-initio} calculations, the quantum-dynamical calculations and the CD calculations are given in the SI Appendices III-VIII.

\section*{Data Availability}

The data presented in this article is available from the corresponding author upon request.

%
%\begin{figure}
%  \includegraphics[width=1.\textwidth,center]{figures-work/first-draft-pdfs/fig4.pdf}
%  \caption{}
%  \label{fig4}
%\end{figure}

%\subsection*{Author Contributions}
%
%HJW proposed the experiments. DB developed the optical setup, acquired and analyzed all data. DB developed the theory and performed all calculations. DB and HJW discussed the results and wrote the manuscript.

%\begin{acknowledgements}
%\section{Acknowledgements}
\acknow{DB thanks Prof.~Frank Neese for helpful discussions regarding the calculation of the distance-dependent magnetic dipole moments. We thank ETH Zurich for the allocation of computational resources on the EULER cluster. We thank Prof.~Ivan Powis (Nottingham) for suggesting to study 2-iodobutane in this work. This work was supported by the Swiss National Science Foundation under project $200021\_172946$ and the NCCR-MUST.}

\showacknow{} % Display the acknowledgments section

%

%\begin{suppinfo}

%Details on data acquisition and analysis; Energy-dependent CD (maximal values and intensity dependence); Results of the time-dependent Schr\"{o}dinger equation calculations used for predicting the effect of magnetic field interactions during the electron's continuum propagation; Details on the calculations of CD induced by electronic dynamics in the cation.

%\end{suppinfo}

%\bibliography{2-iodobutane,attobib}

\end{document}